# Inelastic Electron Scattering at a Single-Beam Structured Light Wave

Sven Ebel[1,*] and Nahid Talebi[1,2,*]

[1]*Institute of Experimental and Applied Physics, Kiel University, 24098 Kiel, Germany*

[2]*Kiel Nano, Surface and Interface Science KiNSIS, Kiel University, 24118 Kiel, Germany*

E-Mail: talebi@physik.uni-kiel.de, ebel@physik.uni-kiel.de

**Abstract** − In this work we demonstrate the inelastic scattering of slow-electron wavepackets at a propagating Hermite-Gaussian light beam. The pulsed Hermite-Gaussian beam thereby forms a ponderomotive potential for the electron with large enough momentum components, leading to the inelastic scattering of electrons and their bunching along the longitudinal direction. We show that the resulting energy-gain spectra after the interaction is strongly influenced by the self-interference of the electron in this ponderomotive potential. It is shown that this effect is observable for various optical wavelengths and intensities and further discuss how the variation of the electron velocity and the light intensity allow to control the energy modulation of the electron wavepacket. This effect opens up a new platform for manipulating the electron wavepacket by utilizing the vast landscape of structured electromagnetic fields.

The scattering of electrons [1-3], atoms [4], and molecules [5] at light has been the subject of intensive research activities for many decades. All these works are based on the prediction made by Kapitza and Dirac [1] that the matter wave is diffracted from a grating that is formed by two counterpropagating light waves. The thereby acting physics can be understood equivalently either from a particle or matter-wave point of view [6]. In the particle picture, a photon is absorbed by the electron and a second photon is emitted simultaneously via a stimulated process. The required momentum conservation leads to a change in the transverse momentum distribution of the electron. The matter-wave interaction in this case is mediated by the ponderomotive potential forming from the optical standing wave. In the classical force picture, the resulting ponderomotive force pushes the electron out of the high-intensity region leading to a change in the transverse momentum distribution of the electron beam. The experimental realization of the Kapitza-Dirac effect, nearly 70 years after its theoretical prediction [2,3], and the interest in the quantum-coherent control of the electron wavepacket in ultrafast electron microscopes [7-13] lead to a generalization of the Kapitza-Dirac effect in various scenarios. These include the expansion to the multiphoton regime of intense laser fields [14], including relativistic corrections [15], generalization to include two different wavelengths for light beams [16], and the demonstration of quantum-path interference in the Kapitza-Dirac scattering [17]. Beyond this elastic electron-light interaction, there has been an effort to achieve inelastic electron-light interaction in free space. Works in this direction proposed the generalizing of the Kapitza-Dirac effect to laser fields with two frequency components propagating along different directions, thereby realizing inelastic electron scattering from standing bichromatic [18], and traveling bichromatic Fields [10-12,19,20]. The second scheme utilized two inclined laser beams with different wavelengths that causes a dispersive propagation of the electron wavepacket in the potential landscape generated by the optical waves. This kind of interaction was recently demonstrated in two challenging experiments. The first experiment shows the possibility of accelerating the electron beam in vacuum with free-space light [10], while the second experiment demonstrates the modulation of the longitudinal momentum of the electron wavepacket and the formation of an attosecond pulse train [12]. Emerging from the concept of an inelastic Kapitza-Dirac effect, Huang et al. [21] proposed the possibility of creating non-Gaussian matter waves.

Photon-induced near-field electron microscopy (PINEM) is an alternative technique for the energy and momentum transfer between electron wavepackets and light [7-8,22]. PINEM is the result of the inelastic scattering of electrons from optical near fields, where the electrons experience a spectral modulation that shows symmetric quantized sideband peaks. This interaction opened up possibilities for attosecond control of free-electron quantum wavepackets [23] and electron pulse manipulation [24].

Altogether these works tried to overcome the gap in energy-momentum conservation for free-space electron light interaction [22]. So far, all the considered inelastic free-space electron- light interactions are requiring at least two waves at different frequencies and did not consider structured light. In this letter we propose an interaction scheme where a structured monochromatic light wave is used for achieving energy modulation of an electron wavepacket, resulting in a PINEM-like electron spectrum.

We study the horizontal propagation of an electron through a traveling time-harmonic electromagnetic light wave, which is represented by its vector potential $\vec{A}(\vec{r}, t)$. The vector potential in this work is a time-harmonic transverse electromagnetic wave in the shape of a Hermite-Gaussian (HG) beam, a well-known exact solution of the free-space paraxial wave equation in Cartesian coordinates. The considered time-harmonic *x*-polarized HG beam that is propagating along the *y*-axis is given by [25],

$$\vec{A}_{n,m}(x,y,z;\omega) = \hat{x} A_0 \frac{W_0}{W(y)} H_n\left(\frac{\sqrt{2}x}{W(y)}\right) H_m\left(\frac{\sqrt{2}z}{W(y)}\right) \exp\left(-\frac{(x^2+z^2)}{W(y)^2}\right)$$

$$\times \exp\left(i\left[k_{ph}y - (1+n+m)\arctan\frac{y}{y_R} + \frac{k_{ph}(x^2+z^2)}{2R(y)}\right]\right)e^{-i\omega t} \quad (1)$$

Where $\omega = \frac{2\pi c}{\lambda}$ is the angular frequency of the light wave and $A_0$ the amplitude of the vector potential. $W(y) = W_0\sqrt{1+\left(\frac{y}{y_r}\right)^2}$ denotes the in y-direction evolving beam waist with $W_0$ and $y_r = \frac{\pi w_0^2 n_m}{\lambda}$ as the beam waist and Rayleigh range respectively. $R(y) = y\left[1+\frac{y_r^2}{y}\right]$ is the radius of curvature. $H_n$ and $H_m$ are the Hermite polynomials.

In this work we focus on the HG mode of the order $HG_{10}$. Applying this, we obtain the following equations for the vector potentials in two dimensions,

$$A_{1,0}(x,y;\omega) = A_0 \frac{W_0}{W(y)} \frac{2\sqrt{2}x}{W(y)} \exp\left(-\frac{x^2}{W(y)^2}\right) \times \exp\left(i\left[k_{ph}y - (1+n)\arctan\frac{y}{y_R} + \frac{k_{ph}x^2}{2R(y)}\right]\right)e^{-i\omega t}$$

and one dimension $A_{1,0}(x,y=0;\omega) = A_0 \frac{2\sqrt{2}x}{W_0}\exp\left(-\frac{x^2}{W_0^2}\right)e^{-i\omega t}$. We consider an electron wavepacket propagating through such a shaped laser field, experiencing a spatially varying ponderomotive potential along its trajectory on the *x*-axis (Fig. 1). The generation of time-harmonic shaped light pulses has been achieved with a variety of well-established techniques in a broad optical spectral range [26-29]. For understanding the physics of this system, we utilize a previously developed numerical toolbox that solves the time-dependent Schrödinger equation within the minimal-coupling Hamiltonian formalism [13,17,30]. This method allows for retrieving the interaction dynamics by capturing the modulation of the electron wavepacket

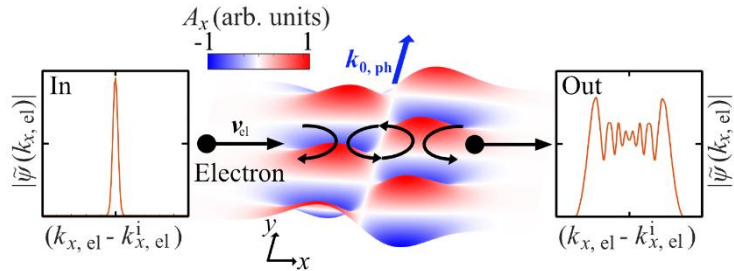

FIG. 1.

Inelastic scattering of a free-electron wavepacket with the center group velocity $v_{el}$ at a traveling Hermite-Gaussian optical beam. The electron is scattered by the resulting Hermite Gaussian shaped ponderomotive potential and experiences a self-interference phenomenon during this process, along the longitudinal direction, that leads to the modulation of the electron kinetic energy after the interaction.

$\psi(\vec{r}, t)$ in the space-time domain. From this we retrieve the wavefunction amplitude in real $|\psi(\vec{r})|$ and momentum $|\tilde{\psi}(\vec{k})|$ spaces at each time step during the interaction. For this we define the transversal momentum $k_{y,el}$ and longitudinal momentum $k_{x,el}$. The electron parameters for controlling the strength of the interaction are the electron initial velocity ($v_{el}$), longitudinal ($W_L$) and transverse ($W_T$) broadening of the electron wavepacket. We first investigate the propagation of a Gaussian electron wavepacket with $W_L = 250$ nm, $W_T = 60$ nm and the center kinetic energy of 1 keV through a time-harmonic HG beam with central wavelength of $\lambda = 700$ nm, temporal broadening of 28 fs and the beam waist of $W_0 = 2\lambda$ (Fig. 2(a)). During the interaction with the HG beam, the phase of the electron wavepacket undergoes a wiggling motion, which is most prominently visualized in the momentum space (Fig. 2(b)). We observe distinct motions along both *x* and *y* directions, i.e., perpendicular and along the propagation direction of the optical beam, respectively. Within the time frame of 0 to 80 fs, the wiggling motion is most pronouncedly directed towards the positive and negative x axis. This oscillating motion of the

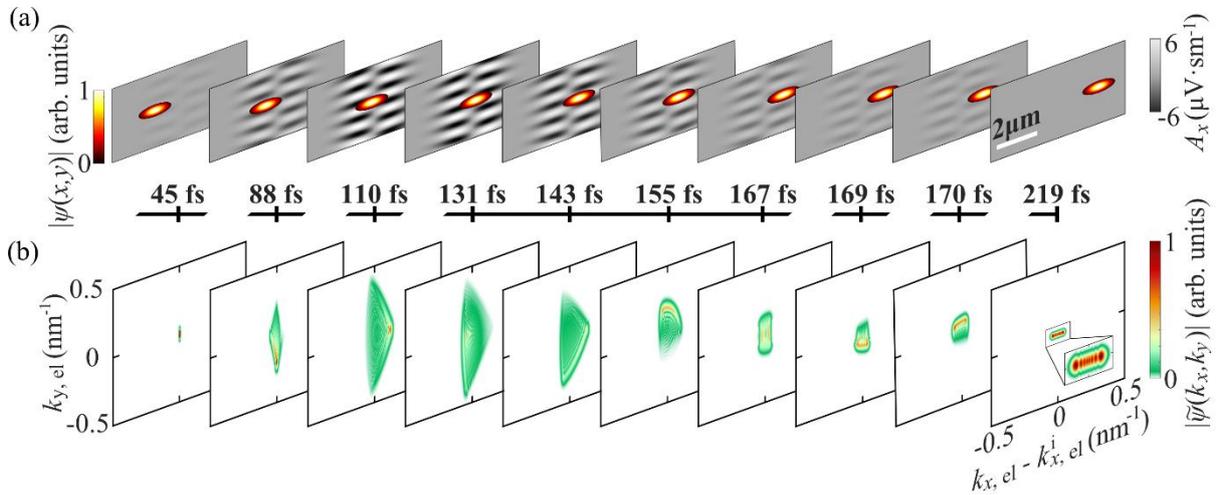

FIG. 2.

Dynamics of the evolution of a Gaussian electron wavepacket in the spatial and momentum space through a Hermite-Gaussian (HG$_{10}$) pulsed laser beam (the laser electric-field-amplitude, its wavelength, and its temporal broadening are $E_0 = 15 \times 10^9$ Vm$^{-1}$, 700 nm and 28 fs, respectively) at different selected time steps. The electron wavepacket has an initial kinetic energy of 1 keV. The electron wavepacket has initial longitudinal and transverse broadenings of 250 nm and 60 nm, respectively. (a) The *x*-component of the vector potential representing the Hermite-Gaussian structured light field (gray background) at depicted time steps, with the insets demonstrating the amplitude of the electron wavepacket. (b) Electron momentum distribution at the corresponding time steps.

electron beam within this time frame resembles the motion of the electron in a pure $HG_{00}$ Gaussian beam, which allows the electron to occupy the higher order transverse momentum states, only spontaneously [31]. For the case of $HG_{00}$ beam, this transverse phase modulation though is averaged out after the interaction and does not lead to a pure momentum modulation, neither in the longitudinal nor in the transverse direction. However, in the case of the $HG_{10}$ beam considered here, this transverse wiggling motion is subsequently followed by a longitudinal oscillation within the time frame of 80 fs to 140 fs, when the electron travels within the low-intensity region of the optical beam, and thereafter again a wiggling motion along the transverse direction is occurred, until $t = 180$ fs, when the electron leaves the interaction region (See the supplementary movie for a better visualization of the wiggling motion of the electron beam). The final longitudinal momentum spectrum of the electron shows a modulation in the shape of an energy comb. This energy comb reveals distinct sidebands for both positive and negative longitudinal momentums. This is an indicator for both energy loss and gain processes on the electron wavepacket during the interaction. The final transverse momentum remains unchanged during the interaction, as one would expect for a single-beam electron-light scattering experiment [31]. Thus, the intermediate time steps visualize the dynamics of the electron populating transversal and longitudinal momentum states, spontaneously. Thereby we observe an oscillation in the momentum state population of the electron. This oscillation alternates between transversal and longitudinal momentum state population. The remained populated momentum orders appear as a pattern of thin maximum and minimum fringes that reassemble an interference pattern. The final longitudinal momentum comb leads to a final energy modulation of the electron wavepacket and reassembles the PINEM spectrum observed for the interaction of electron wavepackets with the near-field light distributions and therefore showing a state population that is similar to a *quantum walk* [32]. The spacing between the observed sidebands cannot be explained through the absorbed and emitted photon energy or momentum recoil, leaving an open question for the physics behind this observed interaction. A known interaction scheme, arising from the ponderomotive force in single beam structured electromagnetic fields, was elaborated by Hilbert et al. [33]. They proposed that the ponderomotive force originating from a Laguerre-Gaussian beam, which can be approximated by a harmonic potential around its origin, acts as a *temporal lens* on the electron, by compressing the wavepacket. This kind of interaction can be regarded as inelastic, but does not lead to the here observed modulation of the electron energy spectrum.

Starting from the observation of the interference phenomena, and that electron-light interactions can be treated as an optical phase modulation of the electron wavefunction within the Volkov approximation [17] we derive here an analytical expression for the estimation of maximum longitudinal momentum gain for the electron in the optical field. The spatial profile of the HG beam in direction of the electron propagation leads to a double barrier ponderomotive potential acting on the passing electron. The observed interference pattern indicates that the electron acquires a phase change induced by the ponderomotive potential creating a Fabry-Perot like geometry for the electron wave – this created system is described visually in Fig. 1. The phase modulation the electron would experience in such a geometry is determined by $\Delta\Phi = 2k_{\text{HG}}d$. Here $d$ is the distance between the reflecting potential planes and $k_{\text{HG}}$ is the maximally supported momentum component by the light interferometer. Further we consider that the phase change that a free electron acquires in an electromagnetic field, in the special case of eikonal approximation is given by the Volkov phase [34] $\Phi_W = \frac{e^2}{2\hbar m_0}\int_0^t \vec{A}(\vec{r},\tau)^2 d\tau$. When simplifying the problem within the 1D $x$-direction, we can insert the vector potential of time-harmonic HG-beam and inserting it in the rearranged expression $k_{HG} = \frac{e^2}{2\hbar m_0}\frac{1}{2d}\int_0^t \vec{A}(\vec{r},\tau)^2 d\tau = \frac{e^2}{2\hbar m_0}\frac{4A_0^2}{W_0^2 v_{el} d}\int_{-\infty}^{\infty} x^2 exp\left(\frac{2x^2}{W_0^2}\right)dx$ we obtain the following expression for the exchanged momentum:

$$k_{HG} = \frac{e^2}{\hbar m_0}\frac{E_0^2}{\omega^2 v_{el}}\frac{\sqrt{\pi}}{4\sqrt{2}}\frac{W_0}{d} \qquad (2)$$

Thereby we used $E_0 = \omega A_0$ and related the time delay $\tau$ to propagation distance $x$ by $x = v_{el}\tau$ with $v_{el}$ being the electron velocity. According to our proposed model we should expect a quadratic dependency from the electrical field strength and an inverse proportionality from the electron velocity for the observed momentum exchange. Therefore, we further numerically studied the proposed system shown in Fig. 1 for different field strengths and electron velocities. Due to the large computational size for the system investigated in Fig. 2, we simplified the calculation by solving the Schrödinger equation for a 1D wavefunction in a 1D external vector potential. The applied algorithm for solving this problem stayed unchanged compared to the 2D solver. To verify these results, we compare the 1D calculation with 2D calculations for selected field strengths

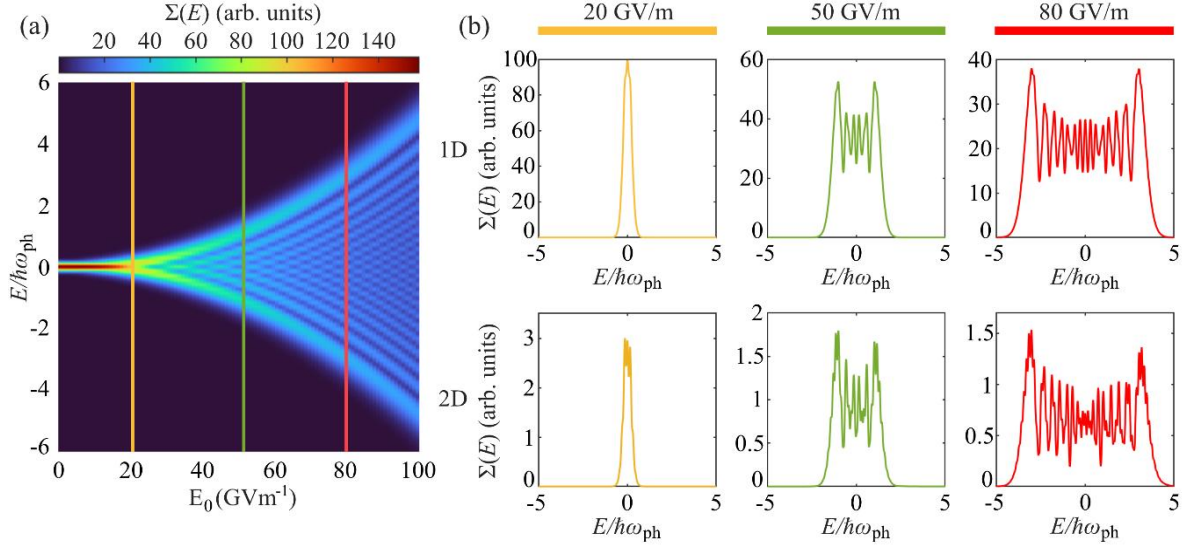

FIG. 3.

The influence of the electric field amplitude on the strength of the inelastic scattering. (a) Electron energy-gain spectrum versus the electric-field amplitude of the incident laser field (the laser wavelength, and its temporal broadening are 200 nm and 8 fs, respectively). The spectra were numerically calculated for a simplified one-dimensional (1D) system. (b) The comparison between 1D and two-dimensional (2D) systems at selected field strength of 20 GV m$^{-1}$, 50 GV m$^{-1}$, and 80 GV m$^{-1}$. The electron wavepacket has an initial center kinetic energy of 1.2 keV. The wavepacket has initial longitudinal and transverse broadenings (2D) of 150 nm and 60 nm, respectively.

and electron velocities. For this we considered a time-harmonic HG beam with a 200 nm wavelength, temporal broadening of 8fs and beam waist of $W_0 = 2\lambda$. The electron wavepacket parameters were adjusted accordingly to $W_L = 150$ nm and $W_T = 60$ nm. The investigated dependence on the electric field strength (Fig. 3) clearly shows an exponential trend leading to an increased energy exchange for strong electric fields. For field strengths below 20 GV m$^{-1}$, there is no modulation of the electron energy spectrum. Field strengths above 20 GV m$^{-1}$ show the sidebands in the electron energy spectrum. These sidebands thereby seem to follow an exponential trend with increased field strength. To review this observed trend, we retrieve the exchanged maximal momentum from Fig. 3(b) at each considered field value. These are then used to fit the expression for the momentum exchange (Equation (2)) (see Supplementary Fig. S1 g) with the distance $d$ as the fit parameter. The resulting fit function ($d_{fit} = 1083$nm) is in good agreement with the numerically calculated maximal momentum exchange. Further noticeable is that the obtained value of $d$, is close to the node to node distance ($d = 900$nm) between the peak amplitudes of the HG profile. The comparison between 1D and 2D calculations further support the observed trends, but indicate that the interaction in 2D leads to additional and more sharp sidebands. This is due to additionally available transverse momentum states in the 2D geometry. When reconsidering the dynamics of the interaction (Fig. 2(b)) we observe a gradual occupation of transverse momentum states. These transverse momentum states open up additional quantum paths for the electron wavepacket. These paths thereby can Interfere with the direct longitudinal transition paths (see Supplementary Material, Movie 1 and Fig. S2), leading to additionally occupied momentum states in the 2D system.

The strength of the momentum exchange versus the electron group velocity shows the expected inverse proportionality (Fig. 4). This means that by decreasing the electron group velocity we observe an increase in the order of the momentum exchange. The enhanced momentum exchange at lower electron energies does not just affect the width of the energy spectrum it also leads to the appearance and sharpening of two distinct sidebands. The 2D simulations again show additional sidebands. The investigation of the velocity dependence further demonstrates the limitations of the proposed phenomenological model here for slow electron velocities. In this specific

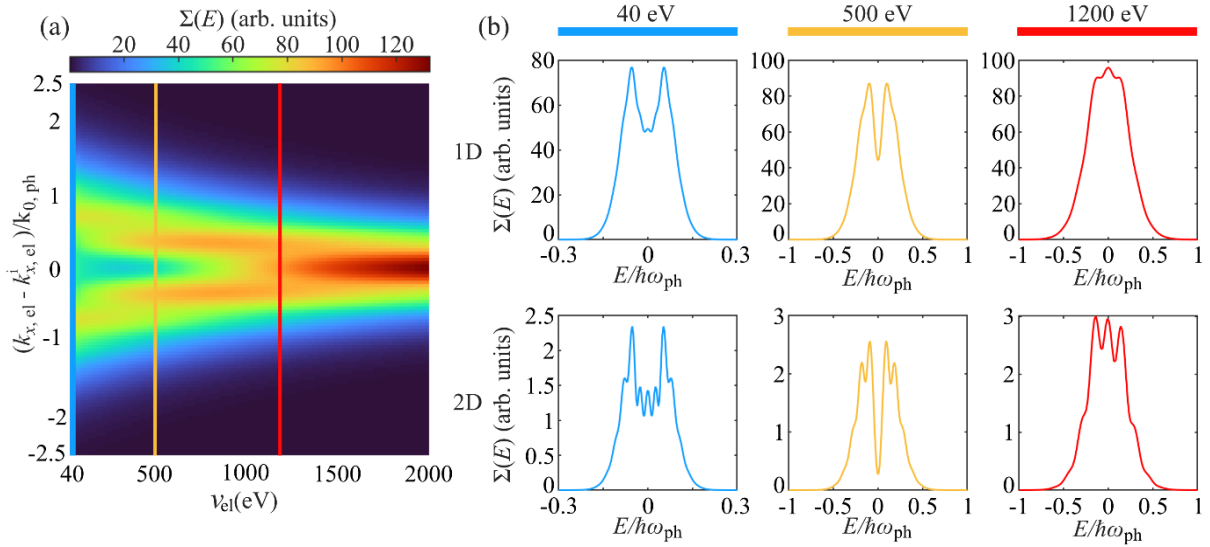

FIG. 4.

The influence of the electron velocity on the strength of the inelastic scattering. (a) The final electron energy-gain spectra versus the electron velocity. The spectra were numerically calculated for a simplified 1D system. (b) The comparison between 1D and 2D systems at selected electron kinetic energies of 40 eV, 500 eV, and 1200 eV. The considered Hermite-Gaussian light field has the electric-field-amplitude, wavelength, and temporal broadening of $E_0=20 \times 10^9$ Vm$^{-1}$, 200 nm and 8 fs, respectively. The wavepacket has initial longitudinal and transverse broadenings (for the 2D case) of 150 nm and 60 nm, respectively.

case we observe the upper interaction limit for electron energies of around 1.5 keV for a laser field with λ = 200 nm and $E_0$ = 20 GV/m. Note that a slow electron also experiences a longer interaction time with the structured electromagnetic field. All together we conclude to have a good agreement between our numerical calculations of the maximum momentum exchange and the analytical estimate given by equation (2).

In summary, our numerical calculations demonstrate a new kind of free-space electron-light interaction. This interaction occurs when an electron moves through a structured time-harmonic laser field. In this work the interaction of an electron wavepacket with a first-order Hermite-Gaussian laser pulse leads to the observation of a PINEM-like electron energy-gain spectrum by showing distinct sidebands. The longitudinal momentum exchange thereby heavily depends on the self-interference of the electron wavepacket, but also interfering quantum paths along the transverse and longitudinal directions. This interaction therefore is substantially different from so far known free-space inelastic electron-light interactions. Our studies further showed that the electron velocity and the strength of the electric field are the key parameters in controlling the described interaction scheme. Furthermore, we were able to calculate the maximum momentum transfer based on a matter-wave Fabry-Perot model. Finally, it should be mentioned, that our results indicate that any structured light beam is feasible for this interaction when being able to support the self-interference of the electron wave-packet in the ponderoomotive potential landscape of the interacting light beam. Feasible candidates might be Laguerre-Gaussian beams, Ince gaussian beams and Bessel beams [35]. This variety of structured light fields as well as the application of this effect together with the Kapitza-Dirac effect might lead to full quantum control of the electron wavepacket via single structured light beams.

**Supplementary Information**

**Outline:**

## S1. Maximum Momentum Exchange

We fit equation (2), from the main text, to the numerically calculated maximum exchanged momentum. In this case we obtain select the data from Fig. 3(a). We obtain a good agreement between the fit function and the data points. The fit parameter d is calculated with $d = 1083$nm, which is close to the node to node distance of 900 nm.

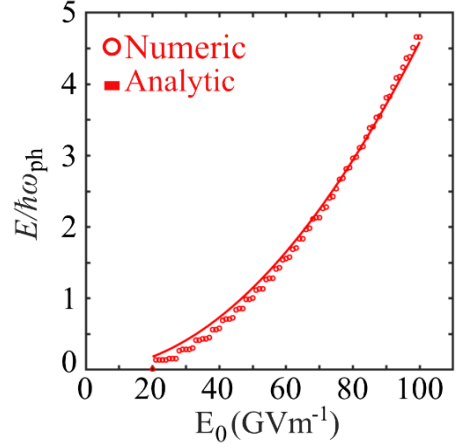

FIG. S1. Fit of the analytical model for the maximal energy exchange to the numerically obtained values where the fitting d parameter is $d = 1083$ nm. The considered electron wavepacket has an initial kinetic energy of 1.2 keV. The wavepacket has initial longitudinal broadening of 150 nm.

## S2. Quantum Paths in the 2D Hermite-Gaussian Beam

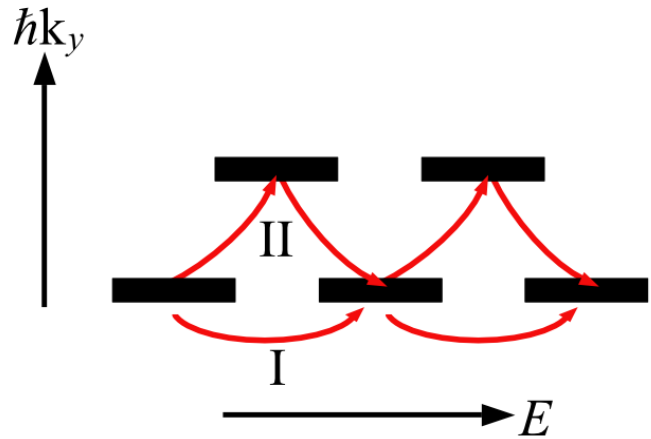

FIG. S2. The schematic of the interfering possible quantum paths in the energy-momentum diagram accessible in the 2D system. The arrows indicate the possible direct energy-state transitions (I) and interfering quantum paths (II).

## S3. Influence of the Wavepacket Dimension

Here we provide a concise additional discussion on the influence of the dimensions of the electron wavepackets on the inelastic scattering of the electron wavepacket at the $HG_{10}$ structured light.

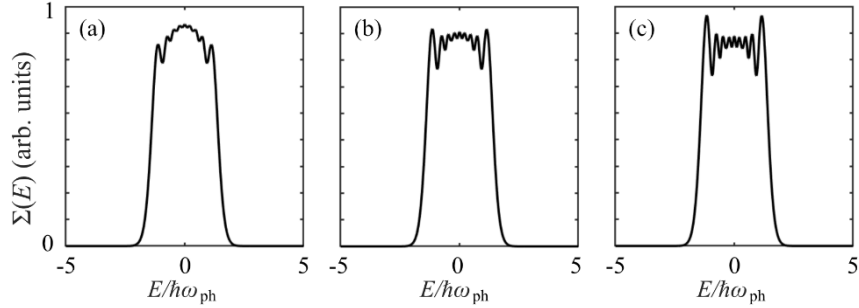

FIG. S3. The influence of the longitudinal broadening of the electron wavepacket on the final electron energy-gain spectra after the interaction. The depicted spectra correspond to longitudinal wavepacket broadenings of (a) $W_L = 130$ nm, (b) $W_L = 140$ nm, and (c) $W_L = 150$ nm. These spectra where numerically calculated for a simplified 1D geometry. The considered Light wave is a Hermite-Gaussian ($HG_{10}$) pulsed laser beam (Laser electric-field-amplitude, wavelength, and temporal broadening are $E_0 = 50 \times 10^9$ Vm$^{-1}$, 200 nm and 8 fs, respectively). The considered electron wavepacket has an initial kinetic energy of 1.2 keV.

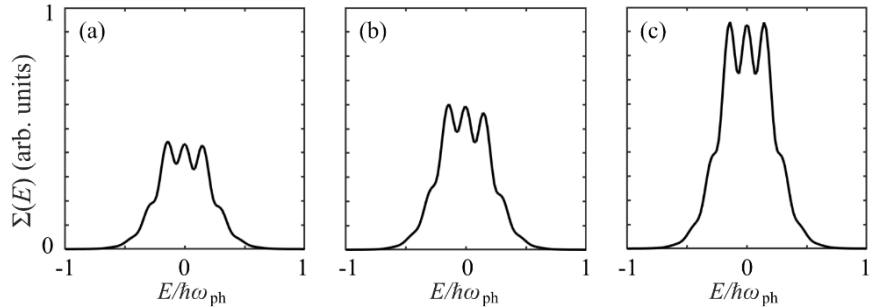

FIG. S4. The influence of the transverse broadening of the electron wavepacket on the final electron energy-gain spectra after the interaction. The depicted spectra correspond to transverse wavepacket broadenings of (a) $W_T = 100$ nm, (b) $W_T = 60$ nm, and (c) $W_T = 20$ nm. These spectra where numerically calculated for a 2D geometry. The considered Light wave is a Hermite-Gaussian ($HG_{10}$) pulsed laser beam (Laser electric-field-amplitude, wavelength, and temporal broadening are E0=$20 \times 10^9$ Vm$^{-1}$, 200 nm and 8 fs, respectively). The considered electron has an initial kinetic energy of 1.2 keV and a longitudinal broadening of $W_L = 150$ nm.

## S4. Explanations for Supplementary Movie

Supplementary Movie "Inelastic Electron Scattering_U1000ev_700nm_50GV" demonstrates the dynamics of an electron wavepacket in a Hermite-Gaussian shaped time-harmonic electromagnetic field. The considered Light wave is a Hermite-Gaussian ($HG_{10}$) pulsed laser beam (Laser electric-field-amplitude, wavelength, and temporal broadening are $E_0 = 50 \times 10^9$ Vm$^{-1}$, 700 nm and 8 fs, respectively). The considered electron wavepacket has an initial kinetic energy of 1.0 keV. Longitudinal and transverse broadening are $W_L = 250$ and $W_T = 60$ nm respectively.